# REDUCED FUNCTIONAL CONNECTIVITY WITHIN THE PRIMARY MOTOR CORTEX OF PATIENTS WITH BRACHIAL PLEXUS INJURY.

*Fraiman D.[4,5], Miranda M.F.[3], Erthal F.[1,2], Buur P.F.,[6] Elschot M.[6], Souza L.[1,2], Rombouts S.A.R.B.[8,9], van Osch M.J.P.[7,10], Schimmelpenninck C.A.[7,11], Norris D.G.[6,12], Malessy M.J.A.[11], Galves A.[3] and Vargas C.D. [1,2]*

[1]*Instituto de Biofísica Carlos Chagas Filho, Universidade Federal do Rio de Janeiro, Rio de Janeiro, Brasil;* [2]*Instituto de Neurologia Deolindo Couto, Universidade Federal do Rio de Janeiro, Rio de Janeiro, Brasil;* [3]*Instituto de Matemática e Estatística, Universidade de São Paulo, São Paulo, Brasil;* [4]*Dept. de Matemática y Ciencias, Universidad de San Andrés, Buenos Aires, Argentina;* [5]*CONICET, Argentina.* [6]*Radboud University Nijmegen, Donders Institute for Brain, Cognition and Behaviour, Centre for Cognitive Neuroimaging, Nijmegen, The Netherlands;* [7]*Leiden University Medical Center, Department of Radiology, Leiden, The Netherlands;* [8]*Leiden Institute for Brain and Cognition, Leiden, The Netherlands;* [9]*Institute of Psychology, Leiden University, Leiden, The Netherlands;* [10]*C. J. Gorter Center for High Field MRI, Department of Radiology, Leiden University Medical Center, Leiden, The Netherlands;* [11]*Leiden University Medical Center, Department of Neurosurgery, Leiden, The Netherlands;* [12]*Erwin L. Hahn Institute for Magnetic Resonance Imaging, University Duisburg-Essen, Essen, Germany*

## Abstract

This study aims at the effects of traumatic brachial plexus lesion with root avulsions (BPA) upon the organization of the primary motor cortex (M1). Nine right-handed patients with a right BPA in whom an intercostal to musculocutaneous (ICN-MC) nerve transfer was performed had post-operative resting state fRMI scanning. The analysis of empirical functional correlations between neighboring voxels revealed faster decay as a function of distance in the M1 region corresponding to the arm in BPA patients as compared to the control group. No differences between the two groups were found in the face area. We also investigated whether such larger decay in patients could be attributed to a gray matter diminution in M1. Structural imaging analysis showed no difference in gray matter density between groups. Our findings suggest that the faster decay in neighboring functional correlations without any gray matter diminution in BPA patients could be related to a reduced activity in intrinsic horizontal connections in M1 responsible by upper limb motor synergies.

Running title: Functional connectivity in brachial plexus lesions

## INTRODUCTION

Brain plasticity consists in the ability of the central nervous system (CNS) to modify in response to changes in behavior, as a consequence of skill acquisition or following central/peripheral injury (Buonomano and Merzenich, 1998; Kaas, 1991; Garraghty and Kaas, 1992). Although a growing body of studies shows that plasticity correlates positively with functional recovery following brain injury (review in Cramer et al., 2011), less is known about the mechanisms underlying functional recovery following peripheral lesion and surgical reconstruction.

Severe traumatic brachial plexus lesions with root avulsion (BPA) leads to motor and sensory function loss of the arm. Although the reconstruction of the original peripheral nerve pathways is not possible, nerve transfer can be performed to regain function. For instance, by connecting the distal denervated musculocutaneous (MC) nerve to the third to sixth thoracic intercostal (IC) nerves (Midha, 2004). Normally the IC nerves are connected to intercostal muscles, which are involved in volitional breathing and postural control. After successful reinnervation of the biceps muscle following intercostal-musculocutaneous (ICN-MC) nerve transfer, the ICN now innervates the biceps muscle. Initially, elbow flexion by biceps contraction can only be effected by respiratory effort, for instance sustained inspiration. In time however, volitional control becomes possible, implying a change in control. Following this surgical procedure, about two-thirds of patients regain biceps function with at least Grade 3 out of 5 according to the Medical Research Council scale (Seddon, 1954; Narakas and Hentz, 1988; Malessy et al., 1993; Malessy et al., 1998; Midha, 2004).

Applying transcranial magnetic stimulation (TMS) to the primary motor cortex (M1), Mano et al. (1995) and Malessy et al. (1998) studied the change in control over the reinnervated biceps muscle some years after ICN-MC transfer performed in patients with BPA. M1 contains a map of movements organized somatotopically (Penfield and Rasmussen, 1950) with gross and largely separated body part subdivisions represented sequentially from lateral to medial precentral gyrus. Mano et al. (1995) and Malessy et al. (1998) found in these operated patients that the biceps representation shifted from medial to a more lateral position in M1 (i.e a shift from the trunk area to the arm area). Although plausible hypotheses have been put forward to understand the role of brain plasticity in recovery of BPA patients after ICN-MC nerve transfer (Mano et al., 2005; Malessy et al., 1988, 2003), it remains unclear which mechanisms underlie the shift from respiratory dependent biceps control to volitional biceps control and to what extent this functional change is a result of plastic changes in the brain.

Some possible, non-excluding mechanisms underlying brain plasticity in M1 of patients with BPA submitted to ICN-MC nerve transfer have been posed so far. Among them, structural plasticity, consisting in new synaptic connections formed between intercostal and biceps areas in M1 after surgical intervention (Malessy et al., 1988, 2003); functional cortical plasticity, in which existent, but previously silent, connections between the two cortical loci in M1 would be strengthened, and plasticity beyond M1, in which brain areas containing motor representations would modulate those of M1. To address to which extent BPA affects functional connections between neighboring regions within M1, we analyze resting-state fMRI and structural MRI data from BPA patients and healthy control subjects.

Resting-state fMRI has already been used to investigate how the human brain's functional organization is affected by BPA (Liu et al, 2013; Qiu et al., 2014). Herein we aim at the effects of BPA on local functional connectivity by exploring the decay of the functional correlations between neighboring voxels within M1. We find evidence that these correlations decay faster as a function of distance in BPA patients as compared to the control group in the M1 region corresponding to the arm but not to the face area. We also investigate whether such larger decay in patients can be attributed to a gray matter diminution in M1 by means of structural imaging analysis. The lack of difference in gray matter density between BPA group and control together with the faster decay in neighboring functional correlations in BPA patients is suggestive of reduced activity in intrinsic functional connections in M1 responsible by upper limb motor synergies.

## MATERIALS AND METHODS

### *Subjects*

Nine right-handed patients with a brachial plexus lesion (mean age 34.6, SD=4.8; mean age at lesion: 18.8 ± 2.2) and eleven right-handed control subjects (mean age 35.4 ± 8 years), matched in age and sex with the patient's group participated in the study. All patients suffered a brachial plexus traction lesion with root avulsion on the right side. They were included in the study only if they had undergone successful Intercostal to Musculocutaneous (ICN-MC) nerve transfer, meaning there was at least some recovery of biceps function (grade of 1 or higher as measured with the Medical Research Council grade).. At the time of the study they showed variable degrees of biceps function recovery (Mean: 3.0, SD 1-4, as measured with the Medical Research Council grade). The exclusion criteria were history of neurological trauma and additional surgical procedures aimed at regaining elbow flexion (e.g. Steindler flexorplasty), and general exclusion criteria for MRI scanning (such as claustrophobia, pacemaker, and metallic implants). The local ethics committee approved the study and the patients gave written informed consent in accordance with the declaration of Helsinki.

### *Experimental procedure*

The volunteers were comfortably positioned inside the scanner during the experiment. Pillows were placed between the forehead of the subject and the coil to minimize head movement. Lower arms were positioned next to the body at a comfortable angle between 10º and 30º by using cushions. The palm of subjects' hands faced up to the extent that this was possible without causing discomfort. Subjects were instructed to keep their eyes closed, and not to think of anything in particular during resting-state scanning. The scanning time lasted for 5 minutes.

*Data acquisition*

To reduce travel time and thereby maximize the willingness of the patients to participate in the study, data were acquired at two centers in the Netherlands. Four patients underwent scanning at Donders Institute (DI) in Nijmegen and five patients, at the Leiden University Medical Center (LUMC). At the Donders Institute (DI) in Nijmegen, measurements were performed on a 3 T TIM Trio MR scanner (Siemens Medical Solutions, Erlangen, Germany). At the Leiden University Medical Center (LUMC), a 3 T Achieva scanner (Philips Medical Systems, Best, The Netherlands) was used. On both systems an eight-channel head coil, which was produced by the same vendor, was used for all data collection. Acquisition parameters were adjusted to be as equal as possible between the two scanners, while still having near optimal settings for each system. The resting state fMRI study was part of a longer protocol that involved also the acquisition of fMRI during a motor task, which is out of the scope of the present paper.

Data were acquired with a single-shot, multi-echo EPI sequence. This sequence allows for maximal functional sensitivity across the whole brain using PAID[35] and retrospective removal of motion artifacts (Buur et al., 2009). At the start of each block, 30 rest volumes were acquired to calculate the weights for PAID echo summation. Five echoes were collected with TE = 9, 23, 36, 50 and 63 ms, using parallel imaging to achieve a threefold increase in data acquisition speed. Online image reconstruction was performed using the GRAPPA[37] and SENSE algorithms[38] on the Siemens and Philips systems, respectively. The whole brain was covered by acquiring 30 axial slices (3.5 mm isotropic voxels, 0.35 mm interslice gap, 64 x 64 matrix). Flip angle = 90o, TR = 2600.

## Data Analysis

### Resting state fMRI data pre-processing

The statistical parametric mapping software package (SPM8, Wellcome Department of Cognitive Neurology, London) was used for part of the preprocessing of resting state fMRI data. The first three functional volumes of each run were removed to eliminate non-equilibrium magnetization effects. The remaining images were corrected for head movement by realigning all the images to the mean image via rigid body transformations. Functional images were co-registered to anatomical images for every subject. Finally, brain images were normalized to standard MNI 152 template using FLIRT (from FSL software, Smith et al., 2004) and data was resampled to 2 mm × 2 mm × 2 mm resolution.

A mask of the Primary Motor Cortex (M1) was taken from Geyer et al. (1996). This mask had MNI coordinates equivalent to those of the functional brain images employed herein. Thus, M1 was defined on each individual's scan. The mask was segmented in five sub-regions (Figure 1A) to compare control subjects and patients coarsely within the body map representation. Since our main interest was to investigate the functional relationship between neighbor voxels, these sub-regions were constructed so that the number of voxels should be roughly the same (figure 1B). Due to the natural irregularity in M1 cortical thickness, this choice led to a different number of slices in the sagittal dimension, mostly for the blue mask (Figure 1C), as well as some variability in the maximum distance between voxels (Figure 1E).

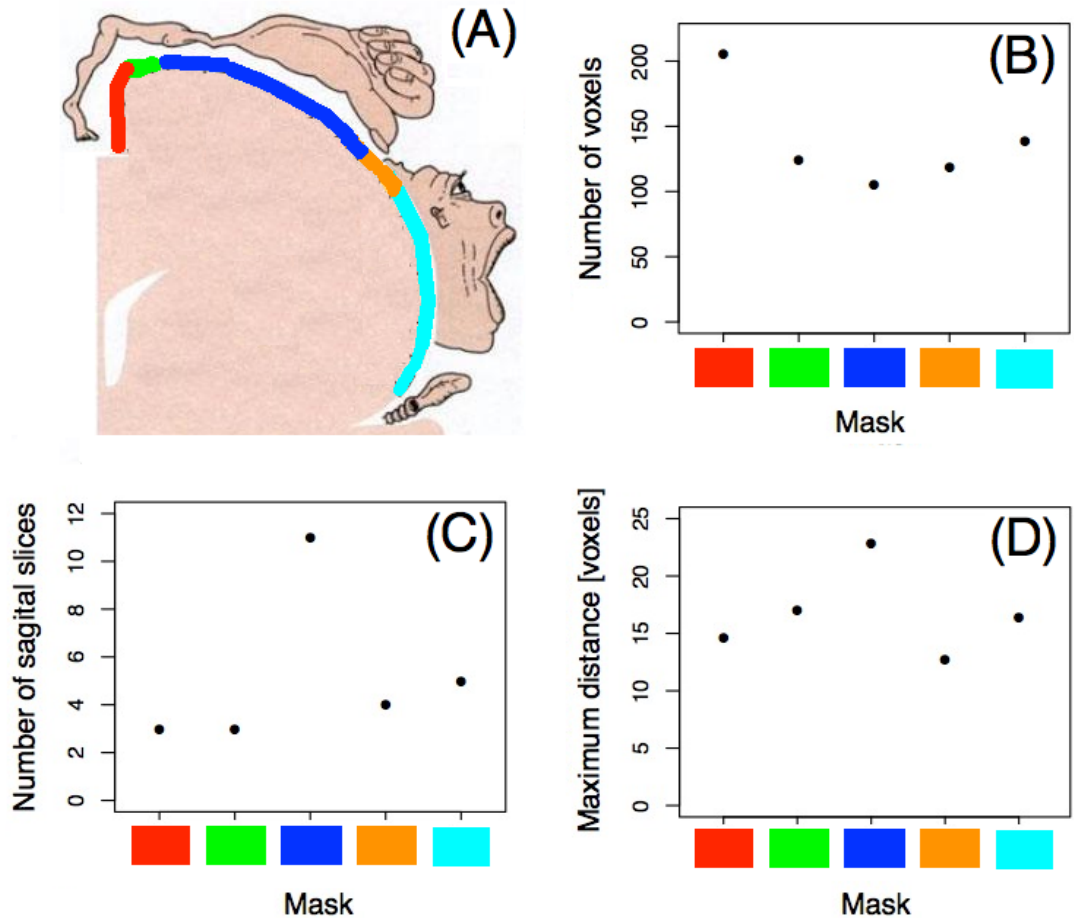

**Figure 1: Masks in the primary motor cortex (M1).** A) Pictorial representation of the human body map in M1 (obtained from http://www.grandesimagenes.com/homunculo/). Each color indicates a mask within M1. B) Number of voxels per mask is roughly equivalent; C) Number of slices in the sagittal plane within the masks; D) Maximum distance between the voxels per mask.

**Resting state fMRI pairwise correlation analysis**

We designed an analysis to investigate local interactions between voxels, to understand how interactions decay as a function of the distance, and most importantly, to compare the correlation behavior between BPA's and controls. Each voxel has two associated quantities: the corresponding resting-state time series and its position (x-y-z coordinates) in the MNI space. The degree of functional interaction between two voxels was calculated using the Spearman (rank) correlation. Here, we considered the correlation to be a function of the Euclidean distance between voxels. Precisely, let $X_t(v)$ represent the value of the resting state BOLD response in voxel v at time t, for $t \in \{1,2,...,T\}$, and let $R_t(v)$ be the ranking of $X_t(v)$ obtained by ordering the sequence from the smallest to the biggest value. Given two voxels v and v', we defined the Spearman correlation between $(X_1(v),...,X_T(v))$ and $(X_1(v'),...X_T(v'))$ as the Pearson correlation between the ranked time series $(R_1(v),...,R_T(v))$ and $(R_1(v'),...R_T(v'))$.

We took the following steps:

First, for each Euclidean distance between voxels, we computed the 95% confidence interval on the population value of the Spearman correlation as aforementioned (Fig. 2 A-E). Second, for each distance we compared the correlations between the two groups, using a Wilcoxon rank-sum test (also known as the Mann-Whitney test) to verify if the two samples come from the same population (Fig. 2 F). Since we are performing one hypothesis tests for each point in distance (total of 70 points of distance that range from 1 to 10) we performed a correction for multiple comparisons.

The corrected threshold was obtained based on the Benjamini-Hochberg procedure (Benjamini and Hochberg, 1995), that controls the false discovery rate $\alpha=0.05$ for the 70 tests performed in each mask. Let us call $H_{(m)}$ the m-th null hypotheses, for m=1,…,70. The procedure consists on the following steps: 1) we order the p-values and let $p_{(k)}$ denote the k-th ordered p-value; 2) for $\alpha=0.05$, we find the largest k such that $p_{(k)} \leq k\alpha/m$, where m is the number of multiple tests (in our case m=70); 3) we reject all H(j) for j=1,…,k.

**Structural MRI:**

The original T1-weighted images of size 198 x 256 x 256 mm3 (voxel size: 1x1x1 mm3) were analyzed. The images were pre-processed using FSL (FMRIB Software Library) and the steps included brain extraction (using BET), linear registration into the MNI152 space (using FLIRT library), and finally the images were segmented into white matter, gray matter and cerebrospinal fluid (using FAST with main MRF parameter equals to zero). The same mask employed to extract pairwise correlations described above (Geyer et al., 1996) was employed to test whether the gray matter tissue on the motor cortex differs among subjects with brachial plexus injury and the control group.

After the pre-processing steps, each voxel on the gray matter tissue map contained a value in the range 0-1 that represents the proportion of that tissue present in that voxel. To identify morphological differences in the brain associated with brachial plexus injury, a Bayesian Spatial Transformation Model (STM) (Miranda et al., 2013) was fitted with the gray matter as the response variables and a covariate vector containing the intercept and diagnostic status (1 for brachial plexus injury and -1 for control). Common voxel-wise methods such as VBM treat voxels as independent units, ignoring important spatial smoothness during the estimation procedure. The STM method used to analyze the structural images is a Bayesian hierarchical model that simultaneously accounts for the varying amount of smoothness across the imaging space and the normality assumption in the model (Miranda et al., 2013).

**RESULTS**

Pairwise Spearman's rank correlation coefficients between voxels computed as a function of voxel distance for each M1 mask in the left hemisphere (contralateral to the affected limb) of controls and BPI patients are depicted in Figure 2. It is clearly evident that higher correlations are found for closer voxels, irrespectively of the region in M1 or the group. However, a closer inspection shows a neat difference among groups (non-overlapping confidence intervals) for the red (Figure 2A), green (Figure 2B) and dark blue (Figure 2C) masks. These more medial regions correspond roughly to the hand and whole body homunculus, as depicted in Figure 1. For these regions, fine grain statistical comparison between groups (Figure 3 and Figure 2F) shows highest differences among controls and BPI patients in pairwise correlations at distances of approximately 8 to 16 mm (4 to 8 voxels). After correction for multiple comparisons, we observed a significant effect in the A, B, and C (red, green, and blue respectively) masks, corresponding roughly to the trunk, upper body and hand areas. Strikingly, no difference between groups is evident in the D and E masks (orange and light blue areas), which corresponds roughly to the neck/face representation in M1.

We also compared the patients with low (n=4, MRC grade 0, 1,2) and high (n=5, MRC 4, see table 1) degree of functional recovery after surgical reconstruction. Our goal was to investigate if the correlations between voxels behaved in a different manner for those two groups. We took the same steps described above to compare controls and BPI's. In each of the 5 masks, we did not find any difference in the correlation behavior between patients with low and high degree of functional recovery.

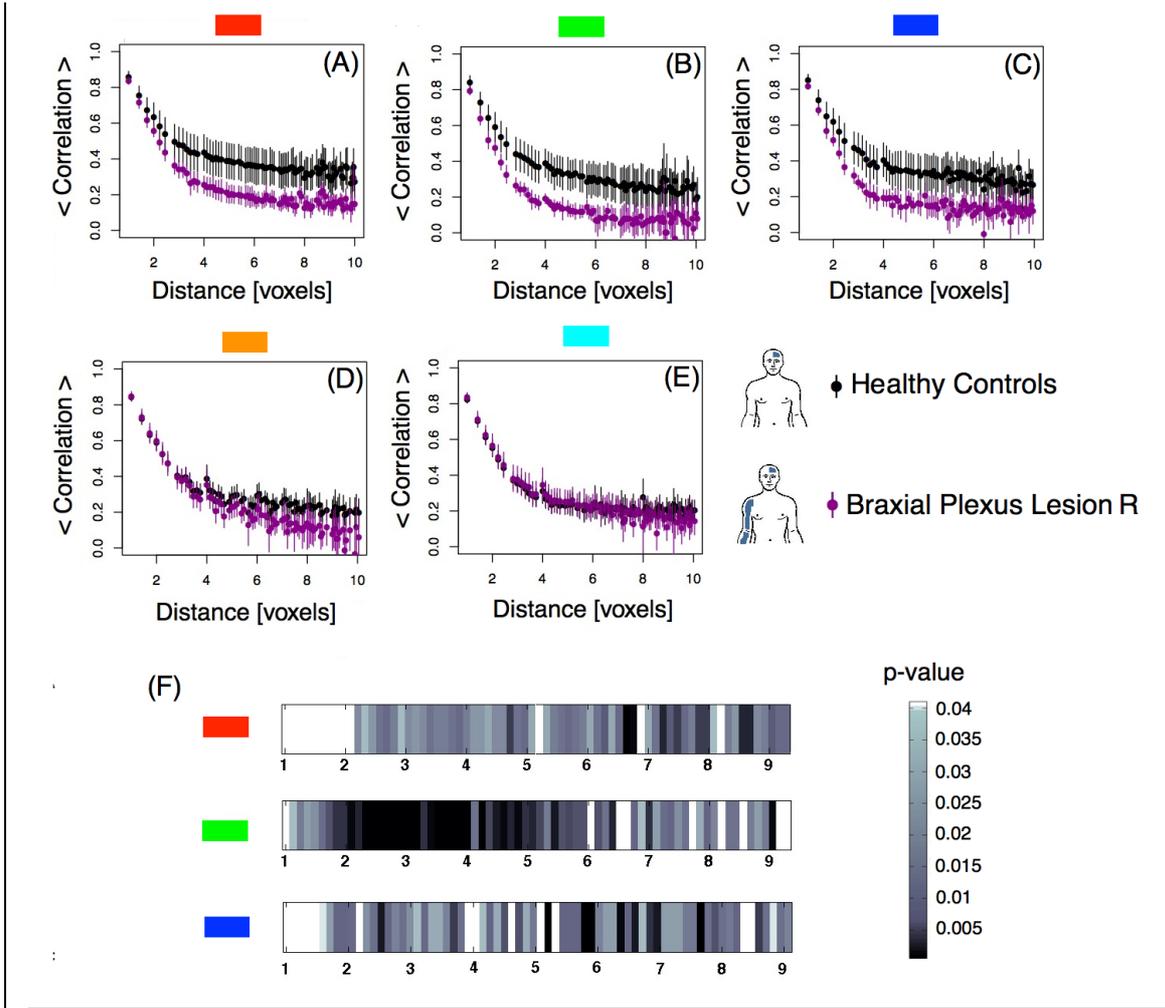

**Figure 2: Pairwise Spearman's rank correlation coefficients between voxels as a function of distance.** A-E: Average correlations per mask as a function of distance between voxels plotted for control subjects and BPI patients. The bars represent the 95% confidence interval. F) Chart representing the level of statistical significance (p-value) resulting from the comparison of the average correlations between groups within masks A, B, and C as a function of distance. Masks D and E did not show any significant difference between the two groups.

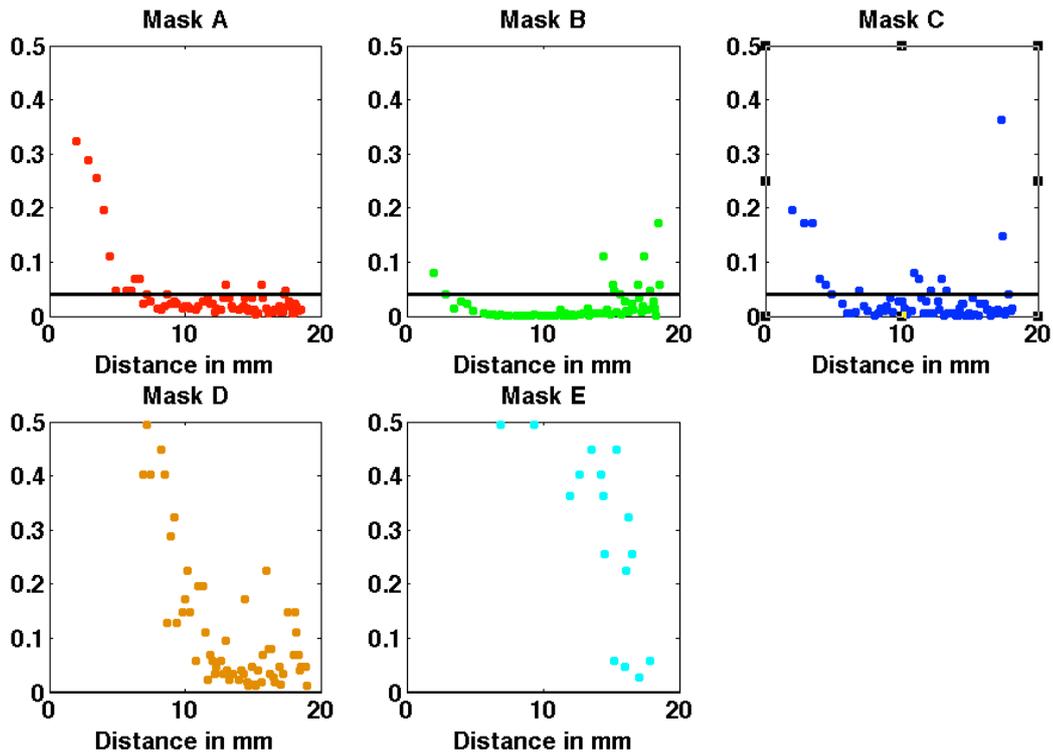

Figure 3: P-values obtained from the Wilcoxon rank-sum test when comparing the correlations of controls versus BPI for each mask A to E. The black horizontal line indicates the p-value threshold, with values below the threshold corresponding to statistical significance after correction for multiple comparisons. The threshold is 0.04 for masks A, B and C and zero for D and E and it was obtained based on the Benjamini-Hochberg procedure (Benjamini and Hochberg, 1995), that controls the false discovery rate alpha=0.05 for the 70 tests performed in each mask.

Figure 4 shows regions of the right and left primary motor cortex (M1) in the sagittal plane with larger differences in functional connectivity between groups. All the regions studied have approximately 100 hundred voxels. For both hemispheres, changes in connectivity were located close to the midline (35 to 45 voxel units) along the sagittal coordinate (Figure 4A). In the left hemisphere (contralateral to the BPA), a second peak was observed more laterally (63 to 65 voxel units) along the sagittal coordinate (Figure 4B). These results suggest that the brachial plexus lesion affects pairwise correlations not only within M1 contralateral to the lesioned limb but also in the more medial portions of the hemisphere contralateral to the spared arm.

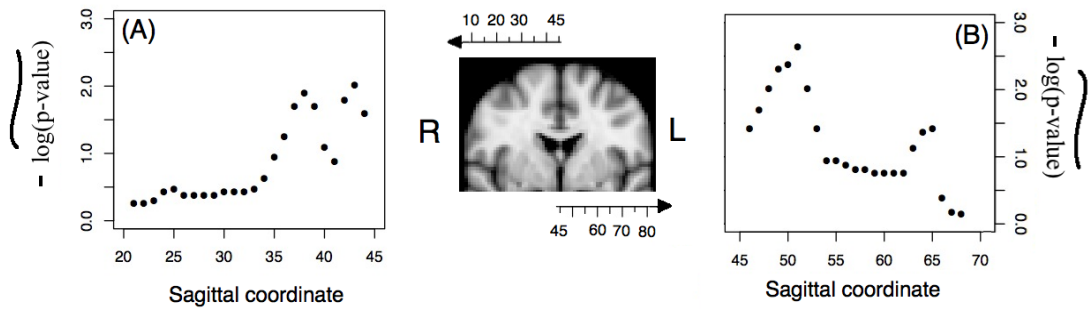

**Figure 4: Changes in functional connectivity in M1 of patients with brachial plexus injury.** The graph represents the statistical difference between control and patient groups for 24 sliding window masks within M1. Each sliding window mask has roughly one hundred nodes, and extended from a given sagittal coordinate (x axis in voxel units) to a second one in the lateral direction. For each mask we computed the correlation function, and tested for equal means for each of the distances using the Wilcoxon test. In the y-axis the median of -log of the p-value (denoted by the tilde superscript) is shown for the hemisphere ipsilateral (A) and contralateral (B) to the right injured arm. R corresponds to the right hemisphere and L, to the left hemisphere. A value above 2 indicates that the median of the set of -log p-values obtained is smaller than 0.01.

To verify if there was any structural difference in M1 between groups, structural imaging analysis was performed using STM (see methods). Results are shown in Figure 5. Inspecting the figure, we notice that for all voxels included in the model, the posterior mean is close to zero. We further computed the 95% credible interval for each voxel and all of them, with no exception, included the value zero. Such a result indicates that there are no differences in gray matter tissue between subjects with brachial plexus injury and the control group.

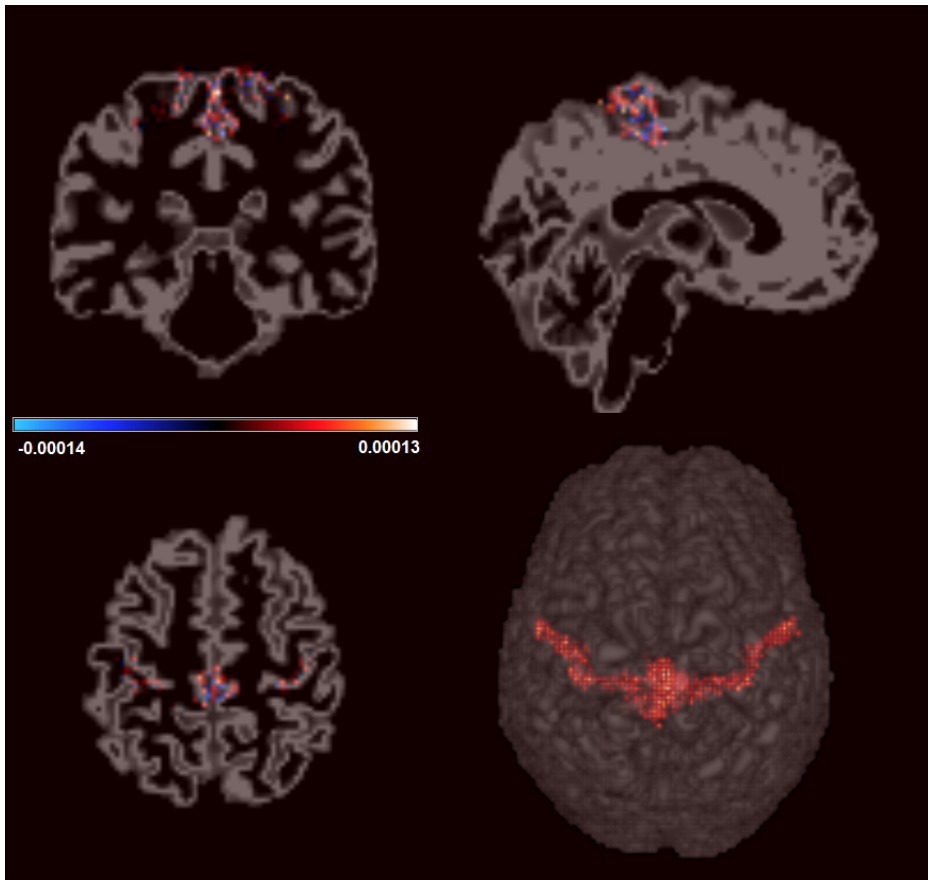

Figure 5- Estimates for the differences in M1 gray matter tissue density between BPI and control group. For each voxel, the color represents the posterior mean estimate of the coefficients associated with the Control versus BPI contrast. As observed in the color bar the values are close to zero indicating that the gray matter density does not differ between groups.

**Discussion**

We herein investigated the correlation decay between neighboring voxels during resting state in the primary motor cortex (M1). We compared control volunteers and in patients that had a traumatic brachial plexus lesion with root avulsions (BPA) in adulthood in whom an intercostal to musculocutaneous (ICN-MC) nerve transfer was performed.

Analyzing correlation as a function of distance between points in space is a classical tool in spatial statistics (Gelfand et al., 2010, Sherman et al., 2011). This tool has already been explored to explore how different regions of the brain interact (see for instance Biswal et al., 1996; Schindler et al., 2007, Lindquist, 2008, Mahmoudi et al., 2012). As expected, higher correlations within M1 were herein found for closer voxels both in controls and in patients. These results are consistent with a modular nature of the cerebral cortex (Douglas and Martin, 2004), where far higher interactions are expected to occur between spatially close units.

From the functional point of view, however, the cortical representation of single movements in M1 seems not to follow a strict modular-based somatotopic pattern, rather obeying neighboring rules which are based on body movements multisegmentar coordination (review in Schieber, 2001). Indeed, mapping studies using intracortical microstimulation (ICMS) in monkeys and functional magnetic resonance (fMRI) in humans revealed small body parts and single muscles in M1 represented in a functionally overlapping, widely distributed mosaic-like fashion (Donoghue et al., 1992; Rao et al., 1995; Sanes et al., 1995). Hence a same body part can be represented at different locations within each one of the gross subdivisions. Transcranial magnetic stimulation (TMS) mapping in M1 further confirm that finger movements evoked over different scalp positions display no prominent topographic pattern (Gentner and Classen, 2006). Furthermore, in monkeys it has been proven that activity in a single corticospinal axon originating from M1 can initiate muscle activity in multiple body parts (McKiernan et al., 1998), suggesting that M1 contains a mosaic representation of motor synergies rather than muscles (Schieber, 2001).

Importantly, horizontal intrinsic connections between spatially distant and functionally different parts of M1 have been consistently revealed in animal models (Huntley and Jones, 1991; Jacobs and Donoghue, 1991; Sanes and Donoghue, 2000; Ziemann, 2004). Thus, whereas the higher functional correlation found in M1 for spatially close voxels could result from intense functional interaction occurring within local modules, lower correlation values observed between voxels at higher distance could result from the concerted activity of long-range horizontal connections within M1. Spanning several millimeters within M1, these long-range horizontal connections were proposed to be involved in activity synchronization beyond cortical modules (Boucsein et al., 2011), fine motor synergy coordination (review in Schieber, 2001) and use-dependent motor learning (review in Sanes and Donoghue, 2000).

Functional connectivity differences were found between healthy controls and BPA injured patients in M1. These differences were most evident in the masks corresponding roughly to the upper limb and trunk representations. However no difference between groups was detected in the mask region corresponding to the face representation, indicating that these changes were specific to body segment representations more directly affected by BPA. A fine grain statistical comparison between groups showed highest differences in pairwise correlations among controls and BPA patients at distances of approximately 8 to 16 mm (4 to 8 voxels). Thus, by strongly disrupting upper limb motor synergies (Schieber, 2001), denervation due to BPA would affect long-range connections and strongly reduce functional connectivity within M1. Reduced pairwise correlations were also verified at more medial regions of the hemisphere contralateral to the spared arm, indicating interhemispheric effects of BPA. These results are in accordance with a resting state functional connectivity approach recently used to explore changes of interhemispheric functional connectivity of motor areas in patients with BPA (Liu et al., 2013).

Pairwise correlation decay as a function of distance was found to occur similarly in patients with low (MRC grade 1 and 2) and high degree (MRC 4) of functional recovery after surgical reconstruction. This lack of difference could indicate that resting state effects are less effective as predictors of functional recovery than, for instance, TMS mapping or a motor task fMRI.

Mano *et al.* (1995) were the first to investigate the location of the cortical area responsible for contraction of the reinnervated biceps muscle in BPA patients, using TMS. Stimulating the motor area situated medially to the cortical area evoking a response in the healthy biceps resulted in a motor evoked potential (MEP) in the reinnervated biceps. At the end stage of recovery, however, the cortical area representing the biceps had shifted laterally in the motor cortex towards the deefferented biceps area (Mano et al., 1995). Malessy et al. (1998) also did not find any difference between the TMS evoked 'cortical location' of the reinnervated and normal biceps areas. Employing fMRI, Malessy et al. (2003) confirmed that the cortical regions in M1 activated during contraction of the surgically reinnervated and the healthy biceps were topologically equivalent. Thus, during recovery, a medial-to-lateral cortical shift might represent a shift from the trunk area to the arm area, possibly underlying the decoupling of volitional breathing from biceps control.

Surgically regaining upper limb function has been proven to reduce pain (Ahmed-Labib et al., 2007, Dimou et al., 2013) and, in the case of ICN-MC transfer, to allow elbow flexion (Narakas and Hentz, 1988; Chuang et al., 1992; Malessy et al., 1998; Venkatramani et al., 2008). Applied to the present study, we hypothesized that successful biceps reinnervation might lead to higher pairwise correlations between the upper limb and the trunk regions. However, our results show a lack of difference between the two groups of patients. This could suggest that long-range horizontal connections do not play a prominent role in functional control following reinnervation. In any case, the severe diminution of the correlation between voxels at intermediate distances observed within M1 could be due to the massive missing of sensorimotor connections with the parts of the arm that have not been surgically reconnected. This in turn would occur in parallel to a greatly reduced repertoire of upper limb synergies (review in Dy et al., 2015).

Although clear functional connectivity differences were found between controls and BPA injured patients in M1, this difference however was not accompanied by gray matter changes. Accordingly, so far peripheral lesions were shown to drive anatomical changes (as measured by gray matter modifications in MRI) in several brain regions such as primary somatosensory cortex, secondary somatosensory cortex, ventrolateral prefrontal cortex, middle cingulate cortex, anterior cingulate cortex and thalamus (Taylor et al., 2009, Davis et al., 2011; Jaggi and Singh, 2011) but not in M1. The reasons for such departure deserve further investigation.

In conclusion, the faster decay in neighboring functional correlations without any gray matter diminution in BPA patients suggests a reduced activity in intrinsic M1 connectivity. This could result from a BPI-induced dysfunction in the horizontal connection intrinsic network, considered to be responsible both of coordinating upper limb motor synergies and driving plastic changes in M1.

Acknowledgement: This work is part of University of São Paulo (USP) project Mathematics, computation, language and the brain, Fundação de amparo a pesquisa do Estado de São Paulo (FAPESP) project NeuroMat (grant 2013/07699-0), CAPES NUFFIC (038/12), Conselho Nacional de Pesquisa (CNPq) (grants 480108/ 2012-9 and 478537/ 2012-3), Fundaçao de amparo a pesquisa do Rio de Janeiro FAPERJ (grants E-26/ 111.655/ 2012 and E-26/ 110.526/ 2012) and Programa de apoyo a la investigacion de la Universidad San Andrés (PAI UdeSA). The funders had no role in study design, data collection and analysis, decision to publish, or preparation of the manuscript.

References

Ahmed-Labib M, Golan JD, Jacques L (2007): Functional outcome of brachial plexus reconstruction after trauma. Neurosurgery 61:1016-23.

Biswal B, Yetkin, FZ Haughton VM, Hyde JH (1995): Functional connectivity in the motor cortex of resting human brain using echo-planar mri. Magnetic Resonance in Medicine34 (4): 537–541,

Boucsein C, Nawrot MP, Schnepel P and Aertsen A (2011): Beyond the Cortical Column: Abundance and Physiology of Horizontal Connections Imply a Strong Role for Inputs from the Surround. Front Neurosci 5: 32.

Buonomano DV, Merzenich MM. Cortical plasticity: from synapses to maps (1998): Annu Rev Neurosci; 21:149-86.

Buur PF, Poser BA, Norris DG (2009): A dual echo approach to removing motion artefacts in fMRI time series. NMR Biomed 22(5):551-60

Chuang DCC, Yeh MC, Wei FC. Intercostal nerve transfer of the musculocutaneous nerve in avulsed brachial plexus injuries (1992): J Hard Surg 17:822-28

Cramer SC, Sur M, Dobkin BH, O'Brien CO, Sanger TD, Trojanowski JQ, et al. (2011): Harnessing neuroplasticity for clinical applications. Brain 134:1591-609.

Davis KD, Taylor KS, Anastakis DJ (2011): Nerve Injury Triggers Changes in the Brain. The Neuroscientist 17:407-22.

Dimou S, Biggs M, Tonkin M, Hickie IB and Lagopoulos J Motor cortex neuroplasticity following brachial plexus transfer (2013): Frontiers in Human neuroscience 7, 500

Dy CJ, Garg R, Lee SK, Tow P, Mancuso CA, Wolfe SW (2015): A systematic review of outcomes reporting for brachial plexus reconstruction. J Hand Surg Am. 40(2):308-13. doi: 10.1016/j.jhsa.2014.10.033. Epub 2014 Dec 13.

Donoghue JP, Leibovic S, Sanes JN (1992): Organization of the forelimb area in squirrel monkey motor cortex: Representation of digit, wrist, and elbow muscles. Exp Brain Res 89(1):1-19.

Garraghty PE, Kaas JH (1992): Dynamic features of sensory and motor maps. Curr Opin Neurobiol 1992;2(4):522-27.

Gelfand AE, Diggle PJ, Fuentes M, Guttorp P (2010): Handbook of Spatial Statistics. CRC Press.

Gentner R, Classen J. (2006):Modular organization of finger movements by the human central nervous system. Neuron 52(4):731-42.

Geyer S, Ledberg A, Schleicher A, Kinomura S, Schormann T, Bürgel U, Klingberg, T, Larsson J, Zilles K, Roland PE (1996):. Two different areas within the primary motor cortex of man. Nature 382:805-807.

Huntley GW, Jones EG (1991): Relationship of intrinsic connections to forelimb movement representations in monkey motor cortex: a correlative anatomic and physiological study. J Neurophysiol 66(2):390-413.

Jacobs KM, Donoghue JP (1991): Reshaping the cortical motor map by unmasking latent intracortical connections. Science 251(4996):944-7.

Jaggi AS, Singh N. Role of different brain areas in peripheral nerve injury-induced neuropathic pain (2011): Brain Research  1381:187-201.

Kaas JH. Plasticity of sensory and motor maps in adult mammals (1991): Ann Rev Neurosci 14:137-67.

Kanamaru A, Homma I, Hara T (1999): Movement related cortical source for elbow flexion in patients with brachial plexus injury after intercostal-musculocutaneous nerve crossing. Neurosci Lett 274:203–206.

Lindquist MA (2008): The statistical analysis of fMRI data. Statistical Science 23(4):439–64.

Liu B, Li T, Tang W-J, Zhang J-H, Sun H-P, Xu W-D, Liu H-Q, Feng X-Y (2013): Changes of inter-hemispheric functional connectivity between motor cortices after brachial plexuses injury: A resting-state FMRI study. Neuroscience 243:33–9.

Mahmoudi A, Takerkart S, Regragui F, Boussaoud D, Brovelli A (212): Multivoxel pattern analysis for fMRI data: A review. Computational and Mathematical Methods in Medicine 2012:1-14.

Malessy MJ, Bakker D, Dekker AJ, Van Duk JG, Thomeer RT (2003): Functional magnetic resonance imaging and control over the biceps muscle after intercostal-musculocutaneous nerve transfer. J Neurosurg 98(2):261-8.

Malessy MJ, Thomeer RT (1998): Evaluation of intercostal to musculocutaneous nerve transfer in reconstructive brachial plexus surgery. J Neurosurg 88(2):266-71.

Malessy MJ, van Dijk JG, Thomeer RT (1993): Respiration-related activity in the biceps brachii muscle after intercostal-musculocutaneous nerve transfer. Clin Neurol Neurosurg 95 Suppl:S95-102.

Mann, H. B.; Whitney, D. R (1947): On a Test of Whether one of Two Random Variables is Stochastically Larger than the Other. Ann. Math. Statist. 18, no. 1, 50--60. .

Mano Y, Nakamuro T, Tamura R, et al. (1995): Central motor reorganization after anastomosis of the musculocutaneous and intercostal nerves following cervical root avulsion. Ann Neurol 38(1):15-20.

Douglas RJ1, Martin KA (2004): Neuronal circuits of the neocortex. Annu Rev Neurosci. 27:419-51.

McKiernan BJ, Marcario JK, Karrer JH, Cheney PD (1998): Corticomotoneuronal postspike effects in shoulder, elbow, wrist, digit, and intrinsic hand muscles during a reach and prehension task. J Neurophysiol 80(4):1961-80.

Midha R. Nerves transfers for severe brachial plexus injuries: a review (2004): Neurosurg Focus 16(5):E5.

Miranda MF, Zhu H, Ibrahim JG (2013): Bayesian spatial transformation models with applications in neuroimaging data. Biometrics 69(4):1074–1083.

Narakas AO, Hentz VR (1988): Neurotization in brachial plexus injuries. Indication and results. Clin Orthop Relat Res 237:43-56.

Penfield W, Rasmussen T. (1950). The cerebral cortex of man: A clinical study of localization of function. New York: Macmillan.

Qiu TM1, Chen L, Mao Y, Wu JS, Tang WJ, Hu SN, Zhou LF, Gu YD (2014): Sensorimotor cortical changes assessed with resting-state fMRI following total brachial plexus root avulsion. J Neurol Neurosurg Psychiatry. 85(1):99-105

Rao SM, Binder JR, Hammeke TA, Bandettini PA, Bobholz JA, Frost JA, Myklebust BM, Jacobson RD, Hyde JS (1995): Somatotopic mapping of the human primary motor cortex with functional magnetic resonance imaging. Neurology 45:919-24.

Sanes JN, Donoghue JP (2000): Plasticity and primary motor cortex. Annu Rev Neurosci 23:393-415.

Sanes JN, Donoghue JP, Thangaraj V, Edelman RR, Warach S (1995): Shared neural substrates controlling hand movements in human motor cortex. Science 268:1775–77.

Seddon HJ, Medical Research C, Nerve Injuries C (1954) Medical Research Council (MRC) Results of nerve suture. London: H.M.S.O.

Schieber MH (2001): Constraints on Somatotopic Organization in the Primary Motor Cortex. J Neurophysiol 86:2125-43.

Schindler K, Leung H, Elger CE, Lehnertz K (2007): Assessing seizure dynamics by analysing the correlation structure of multichannel intracranial EEG. Brain 130:65-77.

Smith SM, Fox PT, Miller KL, Glahn DC, Fox PM, Mackay CE, Filippini N, Watkins KE, Toro R, Laird AR, Beckmann CF (2009): Correspondence of the brain's functional architecture during activation and rest. PNAS ç106(31):13040-45.

Sherman M (2011) Spatial Statistics and Spatio-temporal Data: covariance functions and directional properties. Wiley, West Sussex

Smith SM, Jenkinson M, Woolrich MW, Beckmann CF, Behrens TEJ, Johansen-Berg H, Bannister PR, De Luca M, Drobnjak I, Flitney DE, Niazy R. Saunders J, Vickers J, Zhang Y, De Stefano N, Brady JM, and Matthews PM (2004):. Advances in functional and structural MR image analysis and implementation as FSL. NeuroImage23(S1):208-19
22

Taylor KS, Anastakis DJ, Davis KD (2009): Cutting your nerve changes your brain. Brain 132:3122-33.

Venkatramani H., Bhardwaj P., Faruquee SR, Sabapathy R (2008): Functional outcome of nerve transfer for restoration of shoulder and elbow function in upper brachial plexus injury. Journal of Brachial Plexus and Peripheral Nerve Injury 3:15.

Yoav Benjamini and Yosef Hochberg (1995): Journal of the Royal Statistical Society Series B (Methodological) Vol. 57, No. 1, pp. 289-300.

Ziemann U. TMS induced plasticity in human cortex (2004): Rev Neurosci 2004;15(4):253-66.